# Quantum Mechanical Indeterminacy and Conservation Laws

Moses Fayngold

*Department of Physics, New Jersey Institute of Technology, Newark, NJ 07102*

Conservation laws are discussed in conjunction with quantum-mechanical indeterminacies of the corresponding observables. The considered examples show that the connections between energy and its indeterminacy may be quite intricate. The indeterminacies of the parts of a composite system may be correlated in such a way that the net value of the considered observable is definite. This may be the case with specific entangled superposition of eigenstates in a composite system. In some cases the fact of energy conservation cannot be convincingly established and generally the concept of conserving observable is not universally defined to embrace all possible situations under one rule.

Keywords: energy, conservation, indeterminacy, photon

## 1. Introduction
### Systems with conserving but indeterminate observables

Quantum-mechanical (QM) indeterminacy is fundamentally important characteristic of a system's state, especially if the respective observable is conserved. In some cases the corresponding indeterminacy is also conserved. For instance, while the wave packet of a massive particle is spreading and evolving in a free space, both – its energy and spectrum and thereby indeterminacy remain the same. But generally, the relation between energy, its indeterminacy and conservation is a tricky question. Moreover, the analysis shows that the energy itself turns out to be not consistently defined to embrace all possible cases. A good review of the current situation in this area can be found in [1]. Generally, this is still a domain of lively discussions and controversial views (see, e.g., [2-11])

The presented work analyses some closely related topics. We will use the conventional Copenhagen interpretation of QM. We will analyze some experimentally feasible situations with the commonly accepted definition of energy and its indeterminacy and even there reveal some unresolved or "fuzzy" problems with energy conservation.

The *exact* energy $\mathcal{E}$ of a closed system is an eigenvalue $\mathcal{E}_s$ of the energy operator $\hat{\mathcal{E}} = i\hbar \frac{\partial}{\partial t}$ where $s$ may be an integer or continuous depending on the spectrum of the respective Hamiltonian. A general case is a superposition of eigenstates

$$|\Psi(\mathbf{r})\rangle = \hat{\Gamma} \mathcal{F}(\mathcal{E}_s) |\mathcal{E}_s(\mathbf{r})\rangle d\mathcal{E}_s \qquad (1.1)$$

Here $\hat{\Gamma}$ stands for integration over continuous and summation over discrete range of spectrum, and the eigenstates in question are written in the coordinate representation as functions of **r**.

In a case like (1.1) the energy of the system is defined as the expectation value of all $\varepsilon_s$:

$$\langle \varepsilon \rangle = \hat{\Gamma} |F(\varepsilon_s)|^2 \varepsilon_s \, d\varepsilon_s \qquad (1.2)$$

We will also need the average of the *squared* energy eigenvalues

$$\langle \varepsilon^2 \rangle = \hat{\Gamma} |F(\varepsilon_s)|^2 \varepsilon_s^2 \, d\varepsilon_s \qquad (1.3)$$

This allows one to define the degree of indeterminacy in state (1.1) as the standard deviation

$$\Delta \varepsilon \equiv \sqrt{\langle \varepsilon^2 \rangle - \langle \varepsilon \rangle^2} \qquad (1.4)$$

Now we can formulate the generalized conservation law, say, for energy, in the following way.

Energy of an isolated system as defined by (1.2) must conserve regardless of whether it is the Hamiltonian's eigenstate or an indeterminate (e.g., a Gamow state). In the latter case it conserves statistically, that is, retains its expectation value.

Even though Gamow states are non-stationary, both quantitative characteristics of system's net energy – its expectation value and its standard deviation – must conserve if the system remains isolated. The same must be true for any other conserving observable [12].

Below, we will see how these formulations work in some isolated systems.

### 2. Spin-entangled electron pair

A well-known textbook case using operational procedure to calculate conserving observables and their indeterminacies is a spin-entangled electron pair

$$|\Psi\rangle = a_1 |\uparrow\rangle_1 |\downarrow\rangle_2 + a_2 |\downarrow\rangle_1 |\uparrow\rangle_2, \quad a_1^2 + a_2^2 = 1. \qquad (2.1)$$

Here observable $\varepsilon$ is an electron spin. Denote our electrons as $A$ and $B$. Since each has only two spin eigenstates, the continuous set of amplitudes $F(\varepsilon_s)$ in superposition (1.1) reduces to only two elements with the amplitudes $a_1 \equiv F(\varepsilon_1), \quad a_2 \equiv F(\varepsilon_2)$ determined by (2.1):

$$F(\varepsilon_s) \rightarrow \frac{1}{\varepsilon} \left[ a_1 \delta(\varepsilon - \varepsilon_1) + a_2 \delta(\varepsilon - \varepsilon_2) \right] \qquad (2.2)$$

Let us assign observer Alice with the spin measurement of electron $A$ and Bob with measuring spin of electron $B$. The general protocol (1.1-4) predicts the following results

$$\langle s\rangle_A =\frac{1}{2}\hbar(a_1^2-a_2^2); \quad \langle s\rangle_B =-\frac{1}{2}\hbar(a_1^2-a_2^2), \qquad (2.3)$$

$$\langle s^2\rangle_A =\langle s^2\rangle_B =\frac{1}{4}\hbar^2, \quad \Delta s_A =\Delta s_B =\frac{1}{2}\hbar^2\sqrt{1-(a_1^2-a_2^2)^2}=\hbar a_1 a_2 \neq 0, \qquad (2.4)$$

and the measurement records confirm the predictions. (To simplify the equations, we may henceforth assume some amplitudes real).

As seen from (2.3), the net spin

$$\langle S\rangle =\langle s\rangle_A +\langle s\rangle_B =0 \qquad (2.5)$$

This is an entangled superposition of eigenstates characteristic of a singlet state. And *each electron's spin in this state has non-zero indeterminacy whereas spin of the whole system is completely defined* [12].

### 3. Photon and macroscopic detector

Consider now a single photon with the continuous spectrum making a wave-packet moving along one direction $x$. We consider only one, e.g. electrical "imprint", since both are moving concurrently when in vacuum. Accordingly, we denote it as a $\Psi$-function used for a scalar particle and write (1.1) as

$$\Psi(x,t)=\int F(\omega)e^{i(kx-\omega t)}d\omega, \quad k=\frac{\omega}{c}=\frac{\varepsilon}{\hbar c} \qquad (3.1)$$

The photon is approaching a multi-detector D formed by a layered structure of plates (Fig. 1). Each plate labeled by an integer $j$ absorbs light within a narrow range $\delta\omega_j$ around its characteristic frequency $\omega_j$ with probability $\delta\mathcal{P}_j$ in the framework of the Kramers-Kronig relation [13, 14]. Suppose the sum of all $\delta\omega_j$ embraces the photon's frequency range $\Delta\Omega$. The whole structure can then be used not only for the efficient photon detection but also for studying its spectrum.

Initially the detector has zero momentum, and dropping its rest energy from the equations, we consider the total energy $\mathcal{E}_{In}$ of the system (more accurately, its expectation value) to be in the incident photon:

$$\langle \mathcal{E}\rangle_{In}=\hbar\int|F(\omega)|^2\omega d\omega, \quad P_{In}=\mathcal{E}_{In}/c, \qquad (3.2)$$

where $P_{In}=\mathcal{E}_{In}/c$ is the corresponding momentum.

Using (1.3) we find the average of the *squared* energy and indeterminacy in state (3.1)

$$\langle \mathcal{E}^2 \rangle_{\text{In}} = \hbar^2 \int |\mathcal{F}(\omega)|^2 \omega^2 \, d\omega, \quad \Delta \mathcal{E}_{\text{In}} = \hbar \Delta \Omega = \sqrt{\langle \mathcal{E}^2 \rangle - \langle \mathcal{E} \rangle^2}; \quad \Delta P_{\text{In}} = \frac{\hbar}{c} \Delta \Omega \quad (3.3)$$

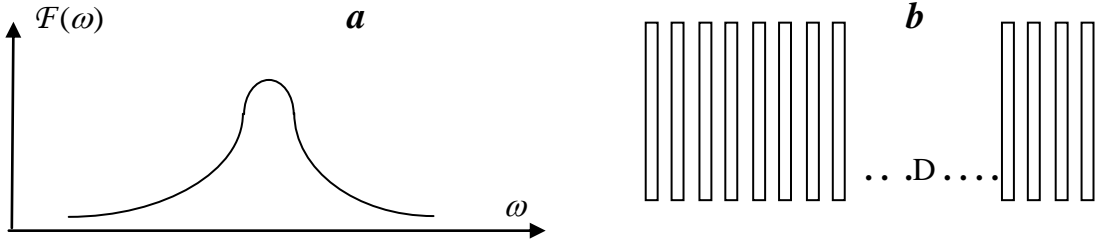

**Fig. 1**
(a) The incident packet and its shape in the $\omega$-basis. (b) The detector D consisting of the system of absorbing plates. Each plate has its specific absorption range and can record the absorption/reflection event.

The photon interaction with D must conserve the net energy, momentum and their respective indeterminacies, but the proof is extremely complicated in view of the complexity of the process.
There remains the above-mentioned possibility to evaluate the initial spectrum of the photon from the statistics of the outcomes including the analysis of post-measurement states of individual plates. It does not clash with the rule that information about the initial state of a measured system is generally irretrievably lost after measurement. Here we must analyze the measurements on a pure ensemble of photons rather than a single measurement. If $\delta\mathcal{P}_j$ are known, sophisticated Math could determine the photon spectrum in Fig.1.
The incident photon faces three basic options: absorption, reflection, and transmission. Ignoring the detector's structure, denote the respective outcomes for the photon as $|0\rangle$ (vacuum state), $|\mathcal{E}\rangle_R$, and $|\mathcal{E}\rangle_T$. The corresponding states of D will be $|D\rangle_0$, $|D\rangle_R$, $|D\rangle_T$. Already on this stage we see a huge number of branches of chains making evolution of the whole system. A coarse-grained description before the system's "collapse" to a final result can be written as

$$|\Phi\rangle = a_0 |0\rangle |D\rangle_0 + a_R |\mathcal{E}\rangle_R |D\rangle_R + a_T |\mathcal{E}\rangle_T |D\rangle_T \quad (3.4)$$

with

$$a_0^2 + a_R^2 + a_T^2 = 1 \quad (3.5)$$

Expression (3.4) is a typical case of an entangled superposition. Calculations of energy and its indeterminacy in such state seems meaningless before separation of $|\mathcal{E}\rangle$ and $|D\rangle$. But after the collapse to a post-measured state the *separate* energy measurements on $|\varepsilon\rangle$ and $|D\rangle$ will show their separate indeterminacies, while the total energy of the system remains sharply defined.

It must also be noted that states R and T in (3.4) and their respective probabilities in (3.5) are results of integration over continuous subsets with various $R(\omega_s)$ and $T(\omega_s)$. The integration will generally produce different shapes of the reflected and transmitted pulse.

As mentioned, calculation of the terms of (3.4) to proof energy conservation is far from trivial. It is even more so when we take account of the physical and geometrical structure of D, e.g., the one shown in Fig.1. Then each of the possible outcomes, e.g. $|\varepsilon\rangle_R$ or $|\varepsilon\rangle_T$, is the result of chain of photon's interactions with the individual plates. For instance, the final state $|0\rangle|D\rangle_0$ may result from successive partial reflections from and absorptions by the individual plates. The probabilities $\delta\mathcal{P}_j$ are themselves the functions of individual properties of each plate.

Each of the three outcomes in (3.4) results from chains of individual interactions. There might be, e.g. the outcome $|\varepsilon\rangle_{RT}|D\rangle_{RT}$ (the same photon partially reflected and partially transmitted, with the corresponding "mixed" state of D). This does not conflict with unitarity. A single electron partially reflected from and partially transmitted through a potential barrier is the same textbook case as a single photon partially reflected from and partially transmitted through a plate. Transition

$$|p\rangle \Rightarrow a_R|p_R\rangle + a_T|p_T\rangle \qquad (3.6)$$

is allowed even for a monochromatic photon, let alone a wave-packet state.

There arises an interesting question about possibility of transition $|p\rangle \Rightarrow |p_R\rangle \otimes |p_T\rangle$. It is forbidden by unitarity for any massive particle but intuitively seems possible for a photon. This is a controversial question deserving a special discussion.

Generally, finding $|\Phi\rangle$ under all possible conditions is a challenging problem especially for a non-monochromatic case in Fig.1. The whole case illustrates the difficulties in tracking the possible changes of system's initial energy and its indeterminacy.

4. **Non-monochromatic photon and an atom**.

Now let us apply the rules formulated in Sec.1 to a system (atom + photon) with the incident photon in a superposition of only two eigenstates

$$|\varepsilon\rangle = a_1|\varepsilon_1\rangle + a_2|\varepsilon_2\rangle \qquad (4.1)$$

Again, assume here, without loss of generality, all probability amplitudes real positive. Here the continuous set $\mathcal{F}(\omega)$ reduces to just two elements $a_1 \equiv \mathcal{F}(\omega_1)$, $a_2 \equiv \mathcal{F}(\omega_2)$ in a superposition mathematically identical to (2.2):

$$\mathcal{F}(\omega) \to \frac{1}{\omega}\left[a_1\delta(\omega-\omega_1) + a_2\delta(\omega-\omega_2)\right], \quad a_1^2 + a_2^2 = 1 \qquad (4.2)$$

The photon frequencies $\omega_1$, $\omega_2$ are tuned to one of the atoms' two optical transitions

$$\hbar\omega_n \equiv \varepsilon_n = E_n - E_0, \quad n = 1, 2 \qquad (4.3)$$

where $E_0$ is the ground level and $E_n$ are the two next energy eigenvalues. We assume otherwise important photonic and atomic line widths negligible relative to $\varepsilon_n$.

The incident photon's energy is indeterminate, with the average

$$\langle \varepsilon \rangle = a_1^2 \, \varepsilon_1 + a_2^2 \, \varepsilon_2 \qquad (4.4)$$

The corresponding indeterminacy value is, according to (1.3)

$$\Delta\varepsilon_{In} = \hbar\Delta\omega = \sqrt{\langle \varepsilon^2 \rangle - \langle \varepsilon \rangle^2} = \sqrt{a_1^2\,\varepsilon_1^2 + a_2^2\,\varepsilon_2^2 - \left(a_1^2\,\varepsilon_1 + a_2^2\,\varepsilon_2\right)^2} = a_1 a_2 \left| \varepsilon_1 - \varepsilon_2 \right| \qquad (4.5)$$

Suppose the atom is initially at rest in its ground state $|E_0\rangle$. But the initial energy of the whole system is indetermined, due to (4.4, 5).

The photon–atom interaction has a variety of possible outcomes.

(IA) Atom gets excited to $E_1$ by absorbing the whole photon:

$$|E_0\rangle \otimes |\varepsilon\rangle \Rightarrow |E_1\rangle \otimes |0\rangle \qquad (4.6)$$

Considering atom as a measuring device, we can say that it has projected the photon's initial state (4.1) onto the initially absent vacuum state $|0\rangle$ (destructive measurement). *But now we cannot say that the energy of the whole system is conserved.* The initial state had, according to (4.4), indeterminate energy with the average

$$\langle E \rangle = E_0 + \langle \varepsilon \rangle = E_0 + a_1^2\,\varepsilon_1 + a_2^2\,\varepsilon_2 \qquad (4.7)$$

and with indeterminacy (4.5). The final state has definite energy $E_1 = E_0 + \varepsilon_1$ different from (4.7). So the total energy of the whole isolated system changes as

$$\begin{Bmatrix} E_0 + a_1^2\,\varepsilon_1 + a_2^2\,\varepsilon_2 \\ \Delta E = a_1 a_2 \left| \varepsilon_2 - \varepsilon_1 \right| \end{Bmatrix} \Rightarrow \begin{Bmatrix} E_0 + \varepsilon_1 \\ \Delta E = 0 \end{Bmatrix} \qquad (4.8)$$

The gained or lost energy is not going to disappear or reemerge even after the atom returns to its ground state, since the energy of the released photon will be definite $\varepsilon_1$ instead of indeterminate according to (4.4).

(IB) The photon is absorbed and the atom gets excited to $E_2$:

$$|E_0\rangle \otimes |\varepsilon\rangle \Rightarrow |E_2\rangle \otimes |0\rangle \quad (4.9)$$

The measurement has again projected the photon's initial state $|\varepsilon\rangle$ onto $|0\rangle$. The outcome is similar to (4.8) with the only distinction of having $\varepsilon_2$ instead of $\varepsilon_1$ on the right.

(IIA) Atom gets excited to $E_1$ and the photon's eigenstate $|\varepsilon_2\rangle$ survives (inelastic scattering):

$$|E_0\rangle \otimes |\varepsilon\rangle \Rightarrow |E_1\rangle \otimes |\varepsilon_2\rangle \quad (4.10)$$

The system's energy changes to

$$E_0 + \langle \varepsilon \rangle = E_0 + a_1^2 \varepsilon_1 + a_2^2 \varepsilon_2 \Rightarrow E_1 + \varepsilon_2 = E_0 + \varepsilon_1 + \varepsilon_2 \quad (4.11)$$

with the difference

$$E_{Fin} - E_{In} = (1 - a_1^2)\varepsilon_1 + (1 - a_2^2)\varepsilon_2 \quad (4.12)$$

The indeterminacy disappears

$$\Delta E_{In} = \Delta \varepsilon_{In} = a_1 a_2 |\varepsilon_1 - \varepsilon_2| \Rightarrow 0 \quad (4.13)$$

(IIB) Similar to the previous case, with labels 1 and 2 swapped.

(III) Atom gets excited to the superposition

$$|E_0\rangle \otimes |\varepsilon\rangle \Rightarrow (a_1 |E_1\rangle + a_2 |E_2\rangle) \otimes |0\rangle, \quad (4.14)$$

This seems to be the most natural outcome: initial superposition of the photon's eigenstates transfers to the same superposition of the atomic eigenstates. It conserves both – the initial energy *and* its indeterminacy.

The whole experiment may be relatively easy to perform. If we consider it as a thought experiment to check the energy and its indeterminacy conservation, I would rather expect the outcome (III) and disprove the predictions (I-II) but not rush to final judgements.

Conclusions to this part:

The considered examples prompt us to think that for an isolated system in a superposition of energy eigenstates the energy expectation value and indeterminacy may generally not conserve. And, by analogy, the same might be true for some other observables like linear and angular momentum. These conclusions contradict the very tenet of the conservation laws. Therefore they ask for critical evaluation.

### 5. Indeterminacy and Lorentz transformations

Consider a photon eigenstate forming a standing wave:

$$|\varepsilon\rangle = \frac{c}{\sqrt{2}}(|p\rangle + |-p\rangle) \qquad (5.1)$$

This is a degenerated state possible when the energy is a function of more than one eigenvalues of a compatible observable like momentum **p** in (5.1). Degeneracy allows $\varepsilon$ to have the same eigenvalue $\varepsilon = \hbar\omega$ for two different momentum eigenstates (the state with eigenvalue $-p$ in (5.1) is *different from state* $-|p\rangle$ with negative probability amplitude). The simplest physical model described by (4.1) may be the lowest mode of the photon with frequency $\omega = \varepsilon/\hbar$ within a Fabri-Perot resonator of length $a = \frac{1}{2}\lambda = \frac{2\pi}{k} = \frac{2\pi c}{\omega}$.

The total energy of the whole system in its rest frame is

$$E = E_0 + \varepsilon = M_0 c^2 + \varepsilon \qquad (5.2)$$

where $M_0$ is resonator's rest mass.

Invite again our observers – Alice to resonator's rest frame K and Bob to a reference frame K′ moving along resonator's axis $x$ with velocity V. In K′, the two magnitudes $|p|$ and the corresponding frequencies are not the same, and degeneracy disappears. The photon's initial eigenstate splits

$$|\varepsilon\rangle \Rightarrow \frac{1}{\sqrt{2}}(|\varepsilon_1\rangle + |\varepsilon_2\rangle) = \frac{c}{\sqrt{2}}(|p_1\rangle + |p_2\rangle) \qquad (5.3)$$

Expression (5.3) is a special case of the continuous superposition (1.1) reduced to two equally-weighted eigenstates with amplitudes given in (3.2) at $a_1 = a_2$.
The individual frequencies and thereby energies of the oppositely directed waves in K′ can be obtained from the Doppler effect as:

$$\varepsilon_1 = \hbar\omega_1 = \hbar\omega\sqrt{\frac{1+\beta}{1-\beta}}; \quad \varepsilon_2 = \hbar\omega_2 = \hbar\omega\sqrt{\frac{1-\beta}{1+\beta}}, \quad \beta \equiv \frac{V}{c} \qquad (5.4)$$

and the total energy of the trapped photon is

$$\langle\varepsilon'\rangle = \frac{1}{2}(\varepsilon_1 + \varepsilon_2) = \frac{1}{2}\hbar\omega\left(\sqrt{\frac{1+\beta}{1-\beta}} + \sqrt{\frac{1-\beta}{1+\beta}}\right) = \varepsilon\gamma(V) \qquad (5.5)$$

Knowing from Alice the values $E_0$ and $\varepsilon$ in K, Bob can obtain the same results directly from the Lorentz transformations. The total energy of the system in K′ is

$$E' = \gamma(V)(E_0 + \varepsilon) \tag{5.6}$$

But since the $\varepsilon'$ in K′ consists of two parts, so is $E'$:

$$E'_1 = \gamma(V)E_0 + \varepsilon_1$$
$$E'_2 = \gamma(V)E_0 + \varepsilon_2 \tag{5.7}$$

The total energy of the system can now be defined as the average of the two eigenvalues observed by Bob:

$$\langle E' \rangle = \gamma(V)E_0 + \frac{1}{2}(\varepsilon_1 + \varepsilon_2) \tag{5.8}$$

It reduces to (5.6) obtained formally by Lorentz transformation from K to K′.
Now, one can prove that indeterminacy depends only on the split parts $\varepsilon_1$, $\varepsilon_2$, so we can write

$$\Delta E = \sqrt{\langle \varepsilon^2 \rangle - \langle \varepsilon \rangle^2} = \frac{1}{2}|\varepsilon_1 - \varepsilon_2| \tag{5.9}$$

This is a special case of (5.5). Note that the state in K′ is not stationary due to energy split (5.7), so its spatial shape must change with time. So the same physical state is stationary for Alice and non-stationary for Bob. This comes apart from the known fact that the resonator is Lorentz-contracted in K′. But the basic result is the emergence of non-zero indeterminacy (5.9). A system with totally definite energy may be in a superposition of energy eigenstates with different eigenvalues in another RF. In other words, indeterminacy *is not a fundamental characteristic.* For the same physical system, a conserving observable may be sharply defined or not, depending on observer.

## 6. Energy conservation and General Relativity

One of the basic effects in general relativity (GR) is "gravitational red shift", for instance, the decrease in frequency of light propagating from region with lower gravitational potential to a region with higher potential. In other words, the frequency of a monochromatic wave is different at different locations [16, 17].

Denoting as $z$ the direction of gravitational force and assuming the field to be uniform, we have for the relative frequency difference $\omega_2 - \omega_1 \equiv \Delta\omega$ between locations $z_1$ and $z_2$

$$\frac{\Delta\omega}{\langle\omega\rangle} \cong -\frac{\Delta\Phi}{c^2}, \tag{6.1}$$

where $\Delta\Phi \equiv \Phi(z_2) - \Phi(z_1)$ is the potential difference between two locations [16 - 18].

This effect was experimentally confirmed in astronomical observations of spectral lines from known chemical elements in the atmospheres of massive stars. In 1959 it was confirmed locally by R. Pound and G. Rebka [19], who compared the proper frequency of the same spectral line at two different levels of the Jefferson tower at Harvard University, using a technique based on the Mossbauer effect [19-20].

The term "gravitational red shift" is not accurate. If the source of light is at a higher potential than our detector, we will observe the blue shift. The mere fact of frequency of monochromatic wave depending on position seems contradiction in terms, but it is normal in GR whose equations are non-linear. Sometimes the gravitational shift is explained in the framework of QM, using the concept of photons. Below we consider two cases – a free photon in gravitational field and a photon in a cavity under gravity.

### 6.1 A free photon in gravitational field

Gravitational shift is the actual frequency shift of a photon as it travels from one place to another. The photon's potential energy at a distance $r$ from the center of a planet of mass $M_0$ is

$$U(r) = m\Phi(r) = -G\frac{M_0 m}{r}, \quad m \equiv \varepsilon/c^2 \qquad (6.2)$$

where $m$ is the relativistic mass of the photon [17, 18]. But this has different incompatible interpretations. If we adhere to QM and consider $\varepsilon = \hbar\omega$ as the universal relation between energy and frequency, then *the energy is not conserved in gravitational field*. In this case, we may, as we did in (6.1), define $\langle\omega\rangle$ between two locations similarly to (4.4) even though the physics and the respective coefficients are quite different.

An alternative is to assume that $\omega$ is associated only with kinetic energy $K$ of the photon, so $\varepsilon = \hbar\omega$ changes to $\varepsilon - U = K = \hbar\omega$. Now frequency change does not affect energy. For instance, if the photon is moving up, its kinetic energy and accordingly, its frequency decrease while $\varepsilon$ does not. Quantitatively, we would have

$$\hbar\Delta\omega = \Delta K = -\Delta U = -m\Delta\Phi \qquad (6.3)$$

which is mathematically identical to (6.1) without affecting $\varepsilon$.

This explanation appeals to our intuition, but one could argue that it is misleading [21, 22]. First, according to QM, the quantity $\hbar\omega$ is an eigenvalue of the photon's Hamiltonian, which includes its *total* energy. The equation $\hbar\omega = K$ holds only in a free space. In a potential field, we have $\hbar\omega = \varepsilon = K + U$, which implies that the photon's frequency must remain constant during its propagation in constant field.

Second, the statement that the photon's frequency depends on detector's location would only make sense if the detector's clock has the same rate at all locations. But this contradicts the GR, according to which the *proper time* of the detector depends on location. So the red shift must result from difference in ticking rate of identical clocks at two different points, rather than from different frequency of the same photon at these points.

One could formulate the gravitational red shift using the Kirchhoff's law, according to which if an atom emits a certain frequency, it absorbs the same frequency. In other words, two identical atoms are ideally tuned to one another.

This law is modified in GR. If a stationary source and detector with identical atoms are held at points with different potentials, they are detuned. This looks like non-conservation of photon's energy. In order to tune them again, we must move them away from or toward each other, depending on their positions. If the detector is at a higher potential, we should start moving it and/or the source toward each other. If the source is at a higher potential, then in order to get them tuned we must move them away from each other. Such formulation avoids any references to energy and correctly describes the whole effect in the operational terms [17].

### 6.2 A trapped photon

Consider first the case without gravity. Interestingly enough, the embryo of the discussed situation starts already in Classical Mechanics in a textbook problem with a non-relativistic particle of mass $M$ trapped in a potential box of width $a$. Its eigenstates are solutions of the Schrodinger equation. The energy eigenvalues in 1D case are

$$\varepsilon_n = \frac{\hbar^2 k_n^2}{2M} = 2\frac{\pi^2 \hbar^2}{M \lambda_n^2} = \frac{\pi^2 \hbar^2}{2Ma^2} n^2, \quad n = 1, 2, 3..., \tag{6.5}$$

where $\lambda_n$ is the wavelength of the corresponding standing wave. In [12] there was obtained an expression for the inner force exerted by the particle on the walls of the box:

$$F_n = \frac{\pi^2 \hbar^2}{M a^3} n^2 \tag{6.6}$$

Now turn to the extreme relativistic case - our photon in a Fabri-Perot resonator of length $H$. The $H$ here plays the same role as $a$ in the potential box. Considering photon as a particle, we can formally attribute to it the mass $m = \frac{\varepsilon}{c^2}$ [17, 18]. A trapped photon is massive (to say the least). Considering it as a wave, we have $m = \frac{\varepsilon}{c^2} = \frac{\hbar \omega}{c^2}$. Now we do not need to solve any wave equation to find the photon energy. It is defined by $H$ from the boundary condition $\lambda = 2H/n$, so we have

$$\varepsilon \to \varepsilon_n = \hbar \omega_n = 2\pi c / \lambda_n = \frac{\pi \hbar c}{H} n \tag{6.7}$$

In contrast with the non-relativistic classical case, where mass is an independent entity, here the photon's mass is determined by the boundary conditions and is quantized. But with all differences, we ask the same question about the inner force exerted by the trapped photon on the resonator's walls. Orient resonator vertically. Changing $H$ requires work $d\varepsilon = FdH$, so

$$F = -\frac{d\varepsilon}{dH} = \frac{\pi \hbar c}{H^2} \tag{6.8}$$

Now turn on the gravity along the optical axis $z$ so $H = z_2 - z_1$.
Even in the simplest case when the field is uniform, the system displays some weird behavior. Frequency $\omega$ becomes function of $z$. Therefore the photon's state is evolving differently than without gravitation. Instead of trivial standing wave or their superposition

$$\Psi(z,t) = \sum_n a_n \sin \omega_n t \cdot \sin \frac{\omega_n}{c} z, \quad a_n = \mathcal{F}(\omega_n) d\omega, \quad \omega_n = \frac{c}{\lambda_n} = \frac{c}{2H} n \tag{6.9}$$

we now have, by virtue of (6.1), each $\omega_n$ changed according to the position $z$ of an observation point as

$$\omega_n \to \omega_n - \omega_n \frac{\Delta \Phi}{c^2} = \omega_n \left(1 - \frac{G \cdot \Delta z}{c^2}\right), \tag{6.10}$$

where $G$ is the generalized "acceleration of free fall" characterizing gravity strength. The state evolution within resonator will now have different time rates at different points. Combining (6.9) and (6.10) gives:

$$\Psi(z,t) = \sum_n a_n \sin \omega_n \left(1 - \frac{G \cdot \Delta z}{c^2}\right) t \cdot \sin \frac{\omega_n}{c} \left(1 - \frac{G \cdot \Delta z}{c^2}\right) z \tag{6.11}$$

Here $\Delta z \equiv z - z'$ is the distance between the chosen point $z$ and the reference point $z'$. The oscillation frequency in (6.11) is higher at the bottom of resonator than at its top, and accordingly the shape loses its synchrony and (anti)symmetry relative to the center.
The local frequency is not explicitly defined. The theory informs only about the *difference* in frequencies between different locations. But since each frequency can be objectively measured at one location, we can surmise that each single eigenstate $\varepsilon_n = \omega_n \hbar$ spreads into continuous range $\Delta \varepsilon_n \approx \left| \hbar \frac{G \cdot H}{c^2} \varepsilon_n \right|$ under gravity. We have to redefine the photon's energy (for each $n$ !) as

$$\varepsilon_n = \hbar \omega_n \to \hbar \langle \omega_n \rangle, \tag{6.12}$$

Any energy eigenvalue loses its status and acquires nonzero indeterminacy.
But there immediately arises the question how to calculate the whole $\langle \omega \rangle$ in a superposition (6.11). Applying general definitions (1.2-3) to considered case, we obtain

$$\langle \varepsilon \rangle = \left(1 - \frac{G \cdot \Delta z}{c^2}\right) \sum_n |a_n|^2 \varepsilon_n, \quad \Delta \varepsilon = \sqrt{\langle \varepsilon^2 \rangle - \langle \varepsilon \rangle^2} \tag{6.13}$$

But consulting with (5.10) we find ourselves again in a vicious circle. The change of all observables is not uniquely defined. According to (5.12), the energy differs by a factor $\left(1 - \frac{G \cdot \Delta z}{c^2}\right)$ when measured at locations separated by $\Delta z$. As in the cosmological observations or Pound-Rebka experiment, where $\Delta z$ is naturally determined by the experimental conditions, here we can evaluate the relevant characteristics using condition $\Delta z = H$. But generally, we have an ambiguity in defining energy and its indeterminacy. Measuring energy according to $\varepsilon = \hbar \omega$, we could only say that the photon's energy is higher at the bottom of the resonator than at its top.

Ironically, at least one manifestation of this ambiguity is quite informative. Unlike the gravity-free case [12], the inner force exerted by the photon on resonator's surface must be different on the top and bottom of resonator. Labeling the corresponding local frequencies as $\omega_B$, $\omega_T$ we can estimate the respective forces $F_B$, $F_T$ as

$$F_B = 2\hbar k_B / \tau_H = \hbar \omega_B / H \quad \text{and} \quad F_T = 2\hbar k_T / \tau_H = \hbar \omega_T / H \qquad (5.13)$$

Here $\tau_H = 2H/c$ is the common period of the photon's complete cycle within resonator. Since $\omega_B > \omega_T$, the force onto the bottom exceeds the one on the top. The difference must be due to the weight $W = mG$ of the photon inside. Indeed, taking the difference $F_B - F_T \equiv \Delta F$ and using (5.1) we obtain

$$\Delta F = \frac{\hbar}{H} \Delta \omega = \frac{\hbar}{H} \langle \omega \rangle \frac{\Delta \Phi}{c^2} = \frac{G}{c^2} \langle \varepsilon \rangle = mG \qquad (5.14)$$

It is just what we had expected. This is an illustration how apparently "fuzzy" definition of photon's energy in GR gives a definite and reasonable prediction when applied under appropriate conditions.

## Conclusions

The concept of energy and its conservation is not universally defined for all theoretically possible situations. It is ambiguous in many cases and may lead to different predictions depending on interpretation. Adding the insult to injury, the concept of energy indeterminacy is not the universal characteristic of a system. Energy of an isolated object may be definite for one observer and indeterminate for another.